\documentclass{sig-alternate-10pt}

\usepackage[hidelinks]{hyperref}
\usepackage{subfig}
\usepackage{tikz}
\usepackage{multirow}

\usepackage{dcolumn}
\newcolumntype{d}[1]{D{.}{\cdot}{#1} }

\title{Bootstrapping Active IPv6 Measurement \\ with IPv4 and Public DNS}
\numberofauthors{1}
\author{
	\alignauthor Stephen D. Strowes\\
	\affaddr{Yahoo Inc., USA}
	\email{sds@yahoo-inc.com}
}

\begin{document}

\maketitle

\begin{abstract}

The IPv4 address space is small enough to allow exhaustive active
measurement, permitting important insight into Internet growth, policy,
and evolution.
The IPv6 address space, on the other hand, presents the problem that we can no
longer perform exhaustive measurements in the same way, inhibiting our
ability to continue studying Internet growth. Access
to private datasets (\emph{e.g.}, HTTP access logs on content servers, flow data in
ISP networks, or passive DNS traces) solves some problems but may not be feasible or desirable.
This paper describes IPv6 address collection by exhaustively sweeping the reverse DNS domain
for the IPv4 address space and performing AAAA queries on the results. Subsequent
ICMP and TCP measurements are conducted to measure the responsiveness of the resulting
set.
%
Key outcomes include: the PTR sweep discovers 965,304 unique, globally routable IPv6
addresses originating from 5,531 ASNs. 56\% of the addresses are responsive, 
across 4,571 ASNs.
Upon inferring pairs of IPv4 and IPv6 addresses that are likely associated with the same device, the data indicates a 
trend toward IPv4 addresses being more responsive than their IPv6 counterparts,
with a higher incidence rate of TCP connections being refused, and wide disparity on where TCP connections or
ICMP echo requests fail silently when comparing IPv4 and IPv6. The disparity in IPv4 and IPv6 responsiveness
is highly variable, and indicative of distinct host configuration and network policies
across the two networks, presenting potential policy or security gaps as the IPv6 network matures.


%
\end{abstract}

\section{Introduction}


Exhaustive active measurement of the Internet's IPv6 address space is infeasible: the
current global allocation defines up
to $2^{125}$ addressable hosts and $2^{61}$ addressable
networks~\cite{iana:ipv6}.
Individual IPv6 BGP advertisements are intractably large for exhaustive scans:
BGP tables from February 29$^{th}$ 2016, 43.5\% of advertisements were
/48s (\emph{i.e.}, $2^{16}$ individual /64s); 25.3\% were /32s.
The pace of growth in the IPv6 network necessitates additional sources
of addresses for direct measurement, or to augment existing heuristics used to
constrain the search space for active measurement.

The problem for active measurement in an IPv6 Internet, therefore, is how to
derive subsets of the IP space to use in measurement studies. There is no
direct address mapping mechanism to determine the IPv6 address of a known-reachable
IPv4 address.
Traffic data is often used to understand the IPv6 address space: HTTP access logs
at content servers, passive DNS traces, or network flow data, all of which can reveal subsets of the active
address space. However, gaining access to these datasets may not be feasible or desirable,
and often the data cannot be released publicly. Public sources will be used to locate
IPv6 hosts for measurement or more nefarious purposes; it is important that the
nature of publicly discoverable devices be understood.

This paper describes a study in which reverse DNS queries are made against all
publicly routable IPv4 addresses advertised on February
$29^{th}$ 2016. The
resulting names are then used to perform forward AAAA queries in order to
derive a large set of around 1 million unique IPv6 addresses.
This technique relies on the assumption that there is often physical and
naming overlap with IPv4 for reasons of minimising costs and simplifying network
management; it is likely to commonly produce IP addresses assigned to infrastructure
in addition to end-hosts and servers.
The types of addresses returned, and the results of active ICMP and TCP
measurements, are presented.
Finally, a conservative pairing of IPv4 and IPv6 addresses is constructed; key to
this part of the study is whether there is disparity, indicating immature
network policy, or differing host configuration.


The contributions of this work include an evaluation of using the IPv4 space
and DNS to discover IPv6 addresses for active measurement, an analysis of
the liveness of those addresses using ICMP and TCP metrics, and an attempt at
pairwise comparison of IPv4 and IPv6 addresses likely assigned to the same device 
to determine whether there are gaps in network or host configuration or policies.


This paper is structured as follows:
Section~\ref{s:background} covers background and related work.
Sections~\ref{s:collection} through~\ref{s:measurement} cover the collection
of PTR records, the collection of AAAA records, then active measurement
against those addresses respectively. 
Section~\ref{s:conclusions} summarises the paper.

\section{Background}
\label{s:background}

Other work has studied aspects of IPv6 deployment and maturity. 
Czyz \emph{et al.} studied various metrics IPv6 deployment metrics~\cite{czyz:2014:ipv6},
including nameservers with IPv6 connectivity, IPv6 glue records,
queries arriving at nameservers, and others including active measurement
on the top 10,000 domains listed in the Alexa
top-million popularity index. The study shows that IPv6 is maturing by
all metrics, though not all at the same pace.

Plonka and Berger contributed traffic measurements from Akamai's content
delivery network, and characterised
the temporal stability of the IPv6 space in~\cite{czyz:2014:ipv6}. Their study
features IPv6 addresses collected from HTTP access logs, and shows how
hosts or ISPs use that space.

On network maturity, Livadaria \emph{et al.}~\cite{livadariu:2016:stability}
compare aspects of IPv4 and IPv6 network stability, including routing stability and
the effect on data-plane stability. Their findings suggest that the IPv6
network is proportionally less stable than the IPv4 network.
Beverly \emph{et al.} also recently studied router availability, albeit without a
contrast to IPv4 stability~\cite{beverly:2015:routeravail}; much earlier work attempted to 
perform global topology discovery~\cite{waddington:2003:topology, kou:2005:topology}

Understanding the structure of the IPv6 Internet, and the means by which anybody
may be able to conduct network measurement or by which attacks may be staged, is 
important; for example, how the IPv6 network structure may be used to launch
malicious traffic~\cite{bellovin:2006:worm}. Other work has listed speculative
approaches or concerns around to active IPv6 measurement~\cite{rfc7707}.

Some work has attempted to use the \emph{ip6.arpa} DNS domain to locate public IPv6
addresses~\cite{van-dijk:ip6.arpa}, though the approach may not return many
responses~\cite{hui:2015:ipv6}. Nikkhah \emph{et al.} also used
the Alexa top-million index to feed DNS queries to locate A and AAAA records to
perform active measurements on connectivity and throughput~\cite{nikkhah:2011:ipv6}.

Earlier work suggested that stateless addressing (SLAAC) or addressing that placed a device's IPv4
address into the bottom 32-bits of its IPv6 were commonplace~\cite{malone:2008:ipv6}. Since then, privacy
extensions for autonomous host addressing were specified~\cite{rfc4941} and have become common primarily
in end-hosts. RFC 7721~\cite{rfc7721} provides an overview of privacy, and address generation mechanisms.


\section{Sweeping \emph{in-addr.arpa}}
\label{s:collection}

The first step of this work is to collect DNS names from which AAAA records
can be queried. The starting point is understanding that, first, the IPv4 space
is currently under high utilisation and, second, that performing
queries to retrieve PTR records for all IPv4 addresses is feasible and cheap.
The \emph{in-addr.arpa} domain is used as a means to perform a query against
an IPv4 address, returning one or more names for a host if the owner of the
address space has provided one.

The principle caveat is to note that there is no operational requirement for
network administrators to configure PTR records for their address space.
While coverage will not be complete, the practice is currently common enough to provide a large set of names.
In many cases, a successful query for a PTR record is likely to
return a name that refers to a single network device.


\subsection{Approach}

To conduct all the DNS queries presented in this section and the active measurements
presented in Section~\ref{s:measurement}, four virtual hosts hosted by
DigitalOcean were deployed in the UK. Each VM runs a local instance of \emph{bind}, 
reachable only by processes on the localhost.

Each host ran a program dedicated to performing the DNS queries
presented in this study (A, AAAA, and PTR), issuing and
handling multiple DNS queries asynchronously with the assistance of
\emph{libevent}.

For each IPv4 address derived from a full routing table collected by Route
Views on February 29$^{th}$ 2016~\cite{routeviews}, a PTR query was sent and
the responses, including error codes, stored for inspection.
This table includes 2,814,910,336 global IPv4 addresses, each of which is used
to create a reverse DNS query for this study.
The AAAA responses are covered in Section~\ref{s:addresses}.

DNS PTR queries took place between March 8$^{th}$ and March 17$^{th}$ 2016.
DNSSEC was not used during this initial study, but should be enabled for future work. In 
cases where the DNS response is too long for a UDP datagram, queries are reissued over TCP.

\subsection{PTR Responses Obtained}
\label{s:method:ptr}

A summary of responses is outlined in Table~\ref{t:ptr-errors}(a). 1.19 billion queries returned names (around 42.8\%).
1.4 billion return an NXDOMAIN error, with no configured subdomain for the
address space. Around 200 million returned a server failure code, and another
3 million where the domain is configured but
no record is found. A small number of queries failed with timeouts.
From the set of 1.4 billion queries that returned an answer, a short breakdown
of obvious misconfigurations of bad data is presented in
Table~\ref{t:ptr-errors}(b).

\begin{figure}
 \includegraphics[width=\columnwidth]{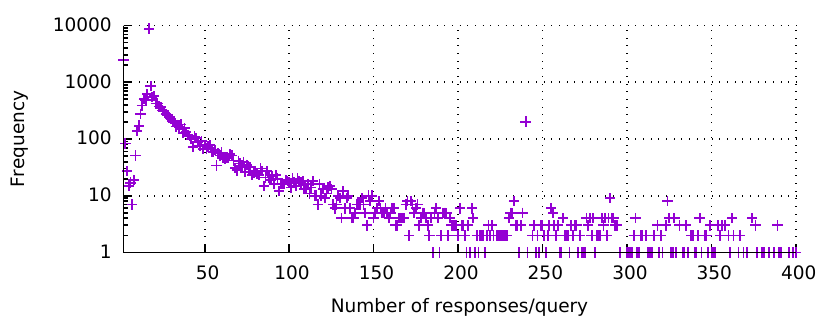}
 \caption{Partial range of PTR response sets larger than one record; full range includes sets with over 1,000 records.}
 \label{f:ptr-response-dist}
\end{figure}

24,945 names returned multiple PTR records. A partial distribution of
the larger answer sets is shown in Figure~\ref{f:ptr-response-dist}, indicating 
that after the common convention of one record per name, the most common PTR sets, though rare, are 15-20 records in size.
The largest set observed from a PTR query was 1,248 records associated to one
IPv4 address. Breaking out the records that returned multiple responses, we have a
full set of 1,190,767,539 names.

The names revealed through this process often identify infrastructure
nodes (routers, middleboxes, firewalls, etc) in addition to hosts intended for use
as public servers.
Common strings surface: ``static'', ``customer'', ``gw''; numeric strings
and two or three-character codes.
The set of names is therefore distinct from datasets based on common web accesses or
DNS requests. In addition, many of the names resolve directly to one network device,
rather than a name that points into a geo-based DNS load balancer.

``localhost'' is associated with 2.4M addresses, and various other strings that appear
to be configuration error are returned; the empty string is associated with almost 1 million
IPv4 addresses, and often the PTR record is a string of the IPv4 address used to create the query.
These are summarised in the right-hand column of
Table~\ref{t:ptr-errors}.

\begin{table}
\small
\subfloat[Response types]{
\begin{tabular}[t]{r|r}
Type & Count \\
\hline
\hline
No domain           & 1,421,766,914  \\
Serv failure        &   199,697,905  \\
No data             &     3,055,458  \\
Other               &        32,112  \\
\hline
No Error            &  1,190,362,811 \\
\hline
Total               &  2,814,915,200 \\
\end{tabular}
}
\subfloat[Notable results]{
\begin{tabular}[t]{r|r}
String & Count \\
\hline
\hline
localhost     & 2,401,398 \\
empty string  &   965,114 \\
IPv4 addr     &   184,858 \\
127.0.0.1     &     1,517 \\
0.0.0.0   &        10 \\
\end{tabular}
}
\caption{PTR query response overview.}
\label{t:ptr-errors}
\end{table}

\section{IPv6 Address Sweep}
\label{s:addresses}

The names discovered in the previous section form the basis for the next stage,
querying for AAAA records.

\subsection{Approach}

Using the same infrastructure as the name collection stage, a AAAA query was
issued for each of the 1,190,767,539 names, and the responses were collected.
These queries were conducted between
March 24$^{th}$ and March 28$^{th}$. Results from the previous section were not de-duplicated,
but \emph{bind} was configured to cache results to minimise the number
of duplicate queries emitted.

%
%
%
%
%

%

\subsection{Results}

\begin{table}
\small
\subfloat[Response types]{
 \begin{tabular}[t]{r|r}
 Type & Count   \\
 \hline
 \hline
 No data        & 856,923,311 \\
 No domain      & 315,419,165 \\
 Server failure &  13,678,831 \\
 Timeout        &       3,414 \\
 \hline
 No error       &   4,742,818 \\
 \hline
 Total          & 1,190,767,539 \\
 \end{tabular}
}
\subfloat[Notable results]{
 \begin{tabular}[t]{r|r}
 Type & Count \\
 \hline
 \hline
 \tt{::1} & 2,401,386 \\
 level3 & 1,024,600 \\
 \tt{prefix::} & 7,715 \\
 \tt{::}  & 84 \\
 \end{tabular}
}
\caption{Overview of results to the AAAA queries.}
\label{t:aaaa-errors}
\end{table}

An overview of the responses in this stage is shown in Table~\ref{t:aaaa-errors}(a); 
4,742,818 queries (0.4\%) returned AAAA records, with particular responses
shown in Table~\ref{t:aaaa-errors}(b): for example, 2.4 million
responses were the result of querying the
``localhost'' string from the PTR sweep, and a further million came from a
common string\\``{\tt unknown.level3.net}'', which all resolved to\\{\tt 2001:1900:2300:2f00::ff} at the time of the study.
This leaves 1,316,832 AAAA responses.
After removal of duplicates and non-routable addresses with no corresponding ASN
in the BGP table, 965,304 unique, globally routable
IPv6 addresses remain. These occupy 328,134 /64s, and originate from 5,531 ASNs.

The number of addresses matching each origin ASN is heavily skewed: 3,132 are small, returning fewer than 10 IP
addresses; 717 IPv6 addresses are the only address from that ASN; 626 ASNs have two IPv6 addresses; 412 have
three. 1,826 return 10 or more, but fewer than 100 IPv6 addresses, and 573 ASNs return
more than 100 IP addresses. The largest of these have tens of thousands of IP addresses in the DNS.


\begin{table}
\small
\centering
\subfloat[ISPs]{
\begin{tabular}{r|r|l}
Network Name & ASN & \#  \\
\hline
\hline
SURFnet (NL)                    & 1103  & 93,649 \\
Deutsche Tele. (DE)             & 3320  & 79,903 \\
1\&1 (DE)                       & 8560  & 51,326 \\
Comcast (US)                    & 7922  & 43,384 \\
GMO (JP)                        & 7506  & 40,672 \\
Yandex (RU)                     & 13238 & 35,109 \\
Host Europe (DE)                & 20773 & 23,522 \\
Hetzner Online (DE)             & 24940 & 19,404 \\
Contabo (DE)                    & 51167 & 15,516 \\
CloudFlare (US)                 & 13335 & 14,387 \\
\end{tabular}
}
\subfloat[Country Codes]{
\begin{tabular}{r|l}
Country & Count \\
\hline
\hline
DE & 237,896 \\
US & 222,528 \\
EU & 123,423 \\
RU & 43,679 \\
GB & 40,082 \\
FR & 33,733 \\
NL & 28,547 \\
CZ & 15,949 \\
NO & 11,355 \\
SG & 11,198  \\
\end{tabular}
}
\caption{Networks and countries with the greatest number of unique IPv6
  addresses discovered.}
\label{t:isps-countries}
\end{table}

Table~\ref{t:isps-countries}(a) enumerates the top 10 ASNs from the data,
given unique occurrences of IP addresses.
The geographic diversity of the networks is evident; notable
is the variety of providers listed: domestic ISPs
are present, but so too are virtual hosting companies, research
networks, and content delivery networks. The range
of networks is of particular importance: passive address collection
from content networks has a tendency to skew towards
collecting address sets from domestic ISPs, and not a
wider variety of networks. Table~\ref{t:isps-countries}(b) indicates the
country-level distribution, according to the country code
for the ASN registered by the regional internet registries.



Table~\ref{t:specialpurpose} describes sets of IPv6 addresses discovered that
match the blocks defined in the IANA special-purpose IPv6 registry.
103,928 queries of those returned one or more of the same \emph{class} of
special-purpose address, outlined in Table~\ref{t:specialpurpose}.

\begin{figure}
%

\subfloat[Frequency of first bit set in IID (note: reverse x-axis)]{
	\includegraphics[width=\columnwidth]{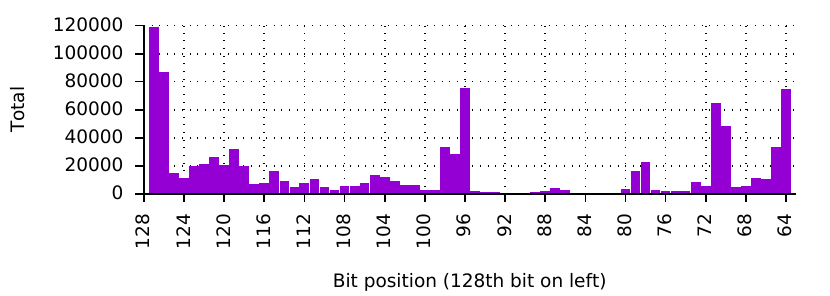}
}

\subfloat[Cumulative form of (a).]{
	\includegraphics[width=\columnwidth]{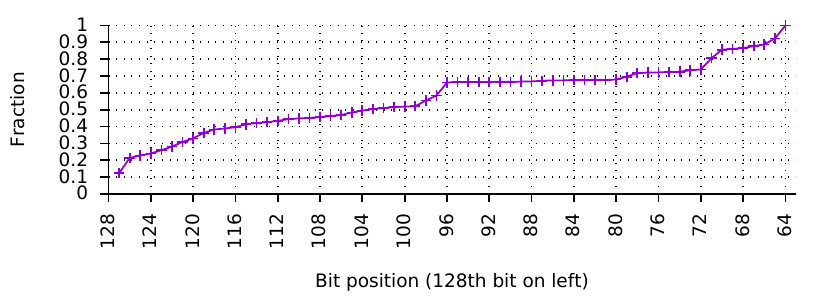}
}
\caption{Bit patterns in the addresses returned.}
\label{f:ptr-bit-patterns}
\end{figure}

Figure~\ref{f:ptr-bit-patterns} indicates structure in the lower 64 bits, the
Interface ID (IID) portion, of the addresses. 
The IID can be generated by various mechanisms; for fixed infrastructure, these are often
statically assigned and likely to follow well-defined patterns.
Using the addresses obtained, these plots show how frequently each bit is the first bit set in the IID portion; that is, they are
a measure of the string of zeros before any set bits. Note that the least significant bit of the address space is on the left-hand
side of the plots. In these addresses, Figure~\ref{f:ptr-bit-patterns}(a) indicates
clearly that the 127th and 128th bits are commonly the only bits set in the addresses
discovered in this study, but there are other bumps near some byte boundaries.
Figure~\ref{f:ptr-bit-patterns}(b) shows the same data in cumulative form. We can
see that for 50\% of these addresses, the first bit set is the 104th bit or
later, and for two thirds of the addresses, the first bit is the 96th bit or
higher. Such indicators are useful in constraining the speculative measurement
space.

\begin{table}
\small
\begin{tabular}{r|r||r|r}
Class & Count & Class & Count \\
\hline
\hline
6to4                     & 109,078 &  ::                             & 84 \\
v4-mapped Addr           & 5,080   &  Teredo                         & 72 \\
::1, not ``localhost''   & 4,350   &  Documentation                  & 18 \\
Link-local               &  981    & IETF      & 5 \\
IPv4-v6 Translat.        &  555    & Direct, AS112  & 3 \\
Unique-Local             &  553    &  & \\
\end{tabular}
\caption{Overview of special-purpose addresses.}
\label{t:specialpurpose}
\end{table}

Other notable address counts are as follows:
\begin{description}
\item [All-zeros:] 7,715 returned addresses have an IID set to zero; in 84 returned addresses, \emph{all} bits are zero.

\item [Stable IIDs:]
73,532 addresses were returned with bits 7 and 25 -- 39 (``\emph{ff:fe}'')
set in the IID, indicating a SLAAC address generated from a MAC address.
72,888, are globally routable, associated with hosts from 1,007 different
ASNs.
RFC 7043~\cite{rfc7043} states that these static  addresses should not be published in the DNS because of the
the privacy concern they present.

\item [Special-Purpose Addresses:]
Table~\ref{t:specialpurpose} lists the number of names that resolve to
special-purpose IPv6 address ranges~\cite{iana:special}. In all, 120,780
queries returned a special-purpose address.
Some 304 names returned \emph{both} globally routable addresses and special-purpose addresses.

\item [Non-standard Addresses]:
28,889 unique responses fall outside the standard IANA allocations. Many of those
are the origin IPv4 address in the bottom 32-bits without the network portion set, much like
the deprecated IPv4-Compatible IPv6 Addresses~\cite{rfc4291}. Many others are 32-bit values
padded into the most signifcant bits of the returned address, which do not
appear to match the origin IPv4 address.

\item [::1:] The origins of the \emph{localhost} strings in
Table~\ref{t:aaaa-errors} is largely
constrained to a small set of ASNs, primarily registered in Vietnam; the
largest contributors are ASNs 45899, 7552, and 7643.

\end{description}

\subsection{Pairing IPv4 and IPv6 Addresses}
\label{s:pairing}

Next, we attempt to consider IPv4 and IPv6 addresses in pairs. Attempting to
pair addresses across protocol families offers scope for direct like-for-like
measurement of performance or network policy, against infrastructure which is
not as heavily monitored as, say, content servers.
It may be possible to determine the gaps in host configuration and security policies
between the IPv4 and the IPv6 networks.

To pair IPv4 and IPv6 addresses, the following approach was used: in each of
the cases where an AAAA query returned at least one result which was a globally
routable address with an ASN in the routing table used throughout this paper,
an additional A query was attempted
on the same domain name. There are of course many cases where names may resolve
to multiple addresses (for either family); to pair addresses, I have
conservatively retained only IP addresses following the following criteria: addresses resolve
from the same name, originate from the same ASN, and where no address resolves
against any other name with a contradictory result.
Note that this could be expanded in cases where the same \emph{organisation} uses multiple ASNs.

This process leads to 673,108 \emph{pairs} of unique IPv4 and IPv6 addresses originating from 5,228 ASNs.

Note there is no requirement that A and PTR records transpose: querying the PTR
record for a given IP address then the A record of the resulting name need not
return the original IPv4 address. Further, the name in the PTR record may have
A and AAAA records, but without any requirement that they map onto the same
network device.
In the common case, however, it is likely that there is a
strong correlation of A records and AAAA records mapping onto a single device
or interface, not least to reduce management complexity.

%

\subsection{Summary}

Although the yield on this form of collection is low, the range of addresses
obtained and the location of those (by ASN, registered country, and
set of ISPs) is broad. As a data source for forming sets of addresses for
measurement, range is important. Broad coverage such as this helps us to understand the subnets actually
in use as a subset of what is visible in BGP, and to better understand
static address allocation patterns within networks. The addresses found here are used in the following section.

\section{Active Measurment}
\label{s:measurement}

In this section, we will use a limited set of active measurements to better
understand how active the set of addresses is. These measurements are intended as a lightweight first attempt at
characterising the addresses, rather than anything that may be construed as
invasive port scanning or flooding.

While tools exist specifically for either port scanning (\emph{nmap})
or rapid measurement of the IPv4 address space
(\emph{zmap}~\cite{durumeric:2013:zmap} and \emph{masscan}),
tooling for large-scale scanning is weaker for IPv6. \emph{scan6} 
attempts to speculatively search IPv6 networks with various heuristics; in
this work, we have a fixed set of target addresses.

The two broad tests conducted on these addresses were ICMP echo requests,
and TCP connection attempts to various well-known port numbers.
These tests used the same virtual machines as used for the DNS queries,
employing GNU Parallel with standard tools: \emph{ping} and \emph{ping6} for
the echo requests, and the standard OpenBSD release of \emph{netcat} for TCP
connections.

The set of unique IPv6 addresses was reordered using the GNU tool \emph{shuf},
and the paired IPv4 addresses identified in Section~\ref{s:pairing} were
inserted adjacent to their IPv6 counterpart. This list was divided across the
four virtual machines, and each individual step scheduled to take approximately
24 hours. The intention here is to be deliberately lightweight, taking
measurements against apparent pairs at approximately the same time but otherwise
attempting to stage the work such that consecutive runs of addresses are avoided.

\subsection{Ethical Considerations}

The measurements for this study were all conducted against IP addresses
publicly listed in DNS: no brute-force or speculative measurement was attempted.
The measurements relied on no unusual TCP or IP options, didn't send any data
aside from packets required for echo requests, TCP handshakes, and TCP teardowns.
TCP connections were closed immediately if successful, and no data from the connection
was read or stored. Nothing other than the outcome of an echo request or a connection
attempt is stored (packets, payload, etc, are all discarded immediately).

The hosts running the measurements also each ran a webserver, configured to serve an informational
page on the study with contact details. 
Five networks requested, via DigitalOcean, that the measurements cease. In
these cases, all addresses advertised from the ASN identified in the report from future measurements.


\subsection{ICMP Echo Requests}
\label{s:method:icmp}

Three ICMP echo requests were sent to each target address, the address marked
active if any responses were received.
Timeouts of 3s were recorded.

\begin{figure}
\includegraphics[width=\columnwidth]{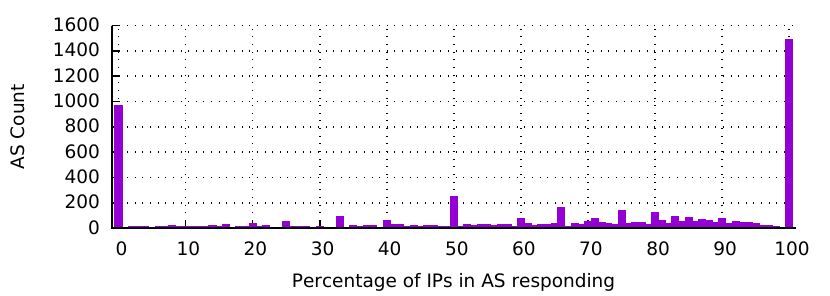}
\caption{Histogram showing distribution of ASNs responding to echo requests.}
\label{f:icmp-hist}
\end{figure}

544,156 addresses (56.7\%) responded to echo requests, across 4,573 ASNs
(82.7\% of the ASNs discovered in Section~\ref{s:addresses}).
The range of responses by ASN, as a percentage of the set of IP addresses per ASN,
is indicated in Figure~\ref{f:icmp-hist}; the clear bimodal pattern is generated
by ASNs with small sets of hosts either all responding or not responding at all, while
the ASNs with larger address sets offer a much more varied response.

\subsection{TCP Measurements}

TCP connections were attempted to the addresses discovered in the previous section,
on a range of port numbers: 21, 22, 53, 80, 443, and 8080.
The purpose is not to be exhaustive, merely to investigate common port numbers
to determine if there are gaps in policy or configuration.
The full list of addresses was tested against one port before progressing to the next.
Connections were staged slowly, paced at around 85,000 addresses an hour.
When a TCP connection timed out (using a 3s timer), no
retry was attempted.

Broadly, on port 80, around 21.1\% of connections were made successfully, and
13.3\% were refused; on port 22, 18.5\% were successful and 9.4\% refused. In
all cases, around 60\% of requests were silently dropped.

Notably, 31,527 addresses accept or reset a TCP connection (on at least one of the ports tested) in cases where
ICMP echo requests were not returned. This may be intentional, if network administrators
have decided to drop ICMP traffic to or from hosts against the current practice specified
in~\cite{rfc4890}.

%
%
%
%
%
%
%

\subsection{Evaluating the IPv4-IPv6 Pairing}
\label{s:analysis}

\begin{figure}[t]
\subfloat[ICMP and TCP response breakdown]{
	\includegraphics[width=\columnwidth]{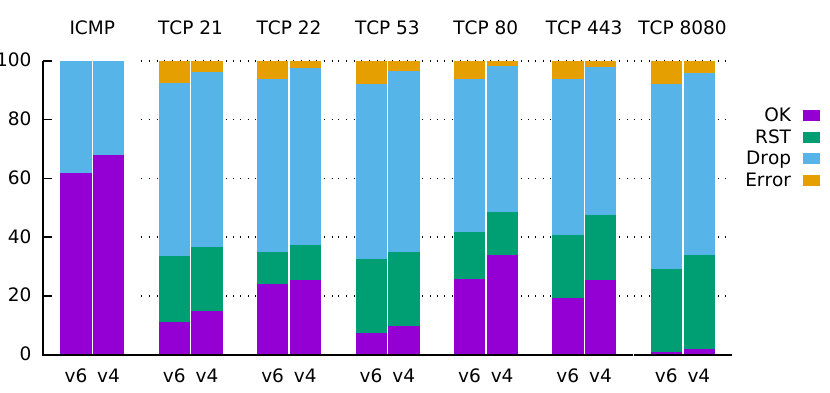}
}

\subfloat[TCP response disparity in ASNs; each subplot is a completion state, each row
is a port number. Quantiles are 5th, 25th, median, 75th, 95th of the spectrum of ASN response rates. Far left
indicates 100\% success on IPv4 and 0\% on IPv6, far right indicates the
opposite.]{
	\includegraphics[width=\columnwidth]{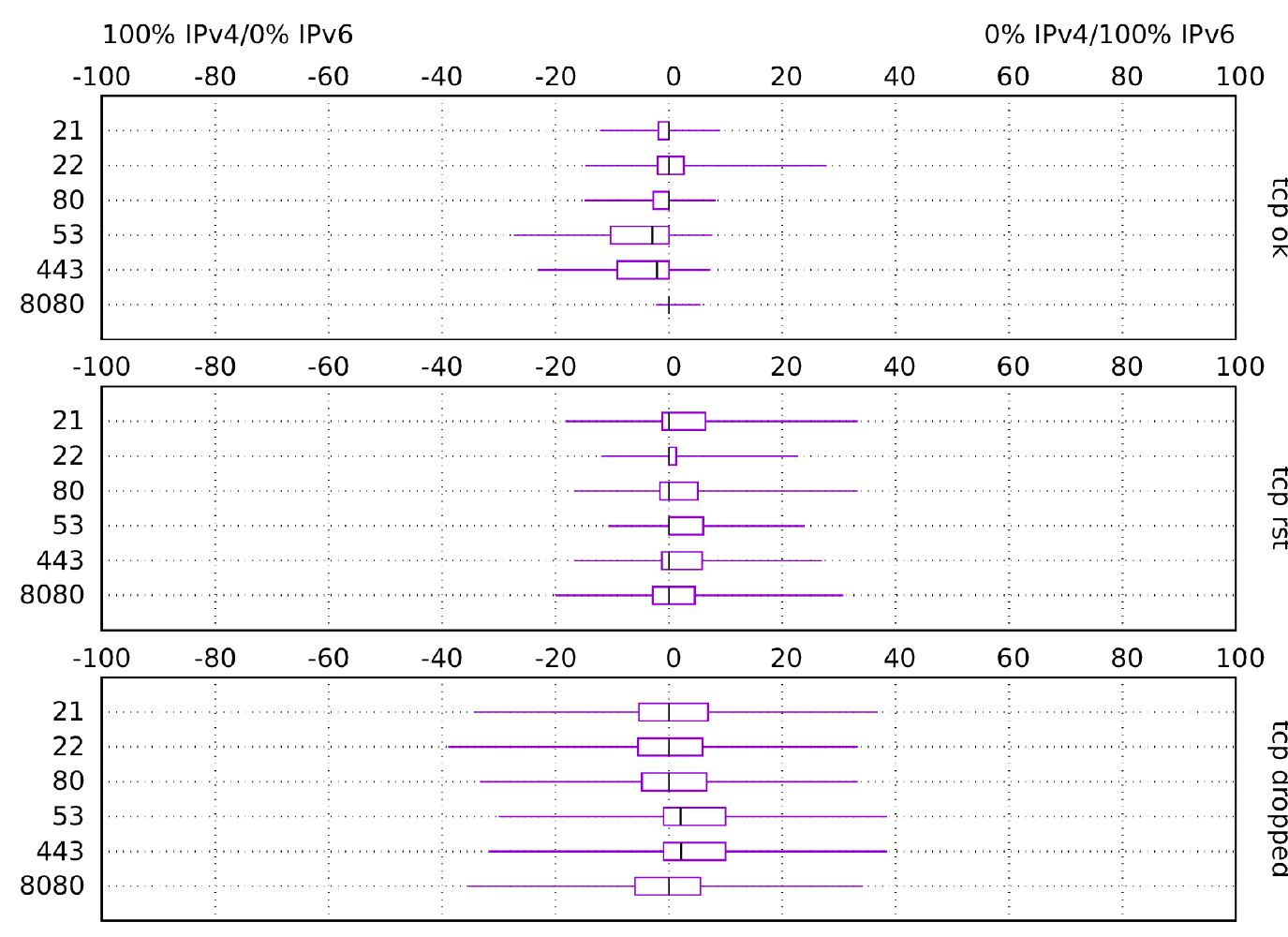}
}

\caption{ICMP and TCP response rates.}
\label{f:tcp-ranges}
\end{figure}

Here, we will refer to the pairs of addresses identified in Section~\ref{s:pairing}.
Figure~\ref{f:tcp-ranges}(a) presents an overview of the response rates for echo requests and
TCP connection attempts on the port numbers tested. First, note that
on this smaller set of addresses, the ICMP echo response rate is slightly higher than
the full set of addresses available, with 61.7\% of the paired IPv6 addresses responding. The
response or success rates on all tests was lower on IPv6 than IPv4, however: 67.8\% of the IPv4
addresses responded to echo requests.

Figure~\ref{f:tcp-ranges}(b) attempts to break down the distributions of the response rates in
each ASN according to the type of the response.
In all cases, IPv4 hosts are more likely to respond to ICMP echo requests,
and are more likely to accept connections on the ports tested, while the IPv6 devices
are more likely to reject connections, implying services are configured to listen on
IPv4 but not yet on IPv6.
TCP drop ratios are, on average, equivalent for IPv4 and IPv6 traffic, though the spread in individual networks is wide.
These trends are not universal, and there is wide disparity in responses across networks.
The data presented here suggests there are gaps in host or network
policy between IPv4 and IPv6; in some cases this may be intentional, for example if server software
isn't deemed stable with IPv6 traffic.

\section{Conclusions/Implications}
\label{s:conclusions}

This paper has presented an approach for discovering IPv6 addresses that can be
used for public measurement. By sweeping PTR records of the IPv4 address space
as advertised in BGP then performing AAAA queries against the returned names,
965,304 unique IPv6 addresses are located in 5,531 autonomous systems.
56.7\% of those respond to ICMP echo requests from 82.7\% of those autonomous
systems.

On attempting to pair IPv4 and IPv6 addresses, 673,108 pairs of unique IPv4
and IPv6 addresses originating from 5,228 are discovered in ASNs. 61.7\% of the IPv6
addresses respond to echo requests compared to 67.8\% of the IPv4 addresses.
Performing TCP measurements against these addresses exposes a similar trend:
devices are more likely to respond over IPv4 than over IPv6. The causes of this
are unclear, but a slightly higher proportion of IPv6 hosts refusing
connections implies host configuration for services lagging network policies.


While this approach relies on the IPv4 space to locate IPv6 addresses on
which measurements can be attempted, the IPv4 network is a resource that
we should not ignore while it is still dominant. Obviously, approaches must
change in future when IPv6 becomes the dominant protocol family.

This study highlights some configuration, privacy, or security concerns.
For example, SLAAC addresses in the DNS may be intentional or accidental.
Similarly, dropped ICMP traffic
may be intentional, or indicative of immature network security policies.

Finally, this study has shown some structure evident in the IPv6 addresses collected.
Bit patterns from real IPv6 deployments, especially for fixed infrastructure,
is useful to help improve existing heuristics for speculative
active measurement studies. Such heuristics are by definition not exhaustive,
but may allow active measurement studies of the IPv6 space similar to the
large body of existing IPv4 active measurement work.

\section{Acknowledgements}

Jonathan Balkind, for reviewing earlier reversions of this paper.

\bibliographystyle{abbrv}
\bibliography{paper,rfc}

\end{document}